\documentclass[9pt,twocolumn,twoside]{osajnl}

\journal{ol} 

\usepackage{bm}

\setboolean{shortarticle}{true}

\title{Simultaneous gain profile design and noise figure prediction for Raman amplifiers using machine learning}

\author[1,*]{Uiara~Celine~de~Moura}
\author[2]{Ann~Margareth~Rosa~Brusin}
\author[2]{Andrea~Carena}
\author[1]{Darko~Zibar}
\author[1]{Francesco~Da~Ros}
\affil[1]{DTU Fotonik, Technical University of Denmark, DK-2800, Kgs. Lyngby, Denmark}
\affil[2]{Dipartimento di Elettronica e Telecomunicazioni (DET), Politecnico di Torino, Corso Duca degli Abruzzi, 24 - 10129, Torino, Italy}
\affil[*]{Corresponding author: uiamo@fotonik.dtu.dk}

\dates{Compiled February 5, 2021}

\begin{abstract} 
A machine learning framework predicting pump powers and noise figure profile for a target distributed Raman amplifier gain profile is experimentally demonstrated. We employ a single-layer neural network to learn the mapping from the gain profiles to the pump powers and noise figures. The obtained results show highly-accurate gain profile designs and noise figure predictions, with a maximum error on average of $\sim$0.3~dB. This framework provides the comprehensive characterization of the Raman amplifier and thus is a valuable tool for predicting the performance of the next-generation optical communication systems, expected to employ Raman amplification.
\end{abstract}

\setboolean{displaycopyright}{true}

\begin{document}

\maketitle

\section{Introduction}
Power efficient erbium-doped fiber amplifiers (EDFAs), conveniently amplifying the C-band, have been applied to optical communication system for more than 25 years . However, their limited and fixed bandwidth operation and noise figures (NFs) higher than 3 dB, make them less attractive to the new generation of high order modulation formats and ultra-wideband transmissions~\cite{Ferrari20}. These limitations have provided renewed interest on the research of Raman amplifiers (RAs) due to three reasons~\cite{Pelouch15}. First, RAs can operate as distributed amplifiers, using the transmission fiber as the gain medium, thus potentially improving the signal-to-noise ratio (SNR) and consequently reducing the effective NF~\cite{agrawalRaman}. Second, they can provide gain at any wavelength, by properly selecting the pump laser frequency. Third, they can broaden the amplified bandwidth when operating in a multi-pump configuration.

Unlike for lumped EDFA-based amplification, in distributed RA the fiber loss is partially counterbalanced by the distributed gain. Therefore, the signal power levels (and consequently the SNR levels) along the transmission fiber can be potentially higher for systems employing distributed rather than lumped amplification~\cite{agrawalRaman}. Assuming an operation in the linear transmission regime, an increase in SNR will directly lead to an increase in the spectral efficiency--transmission distance product. 
Besides providing an increase in the SNR, a key additional advantage of Raman amplifiers is the ability to provide arbitrary gain profiles by adjusting the pump lasers' powers and frequencies. This additional flexibility has a wide--range of potential applications such as: complementing the gain of non--flat amplifiers like EDFAs~\cite{ionescu19}, flattening of frequency combs, and maximizing the throughput in ultra-wideband systems by spectral shaping~\cite{Ferrari20}.

Providing an arbitrary gain profile, in a controlled way, requires selecting the appropriate pump powers and frequencies. We have introduced and experimentally validated a machine learning (ML) framework for determining the corresponding pump powers and frequencies, given a target gain profile~\cite{UiaraJLT_OFC20ext,Zibar20,UiaraJLT20}. This proved to be a versatile tool, able to rapidly provide highly-accurate gain designs for different amplifier schemes such as C~\cite{UiaraJLT_OFC20ext}, C+L~\cite{Zibar20}, and S+C+L-band amplifiers~\cite{UiaraJLT20}, for discrete and distributed configurations. Similar methods were also used to design the RA in hybrid approaches with EDFA~\cite{ionescu19} or semiconductor optical amplifier (SOA)~\cite{Ye2020}, and few-mode RA~\cite{Chen20}. 

So far, the focus has only been on the design of target gain profiles without consider the noise properties~\cite{Zibar20,UiaraJLT20,UiaraJLT_OFC20ext,ionescu19,Ye2020,Chen20}. However, different pump configurations in terms of frequency and power can affect both the signal and the noise power spectral densities during the Raman amplification. Consequently, the flexibility to tune the gain profile by means of pump configuration adjustments can cause undesirable changes in the SNR of the amplified signal, greatly affecting the overall signal transmission performance. Therefore, when designing or adjusting the RA to provide a target gain profile, it is crucial to evaluate the resulting noise profile. 

In this work, the ML framework, based on neural networks (NNs) and presented in~\cite{Zibar20}, is upgraded to incorporate simultaneous prediction of pump powers and NF for a target gain profile. This new framework is now a comprehensive RA inverse design tool, able not only to provide the desired gain profile, but also its noise performance. The experimental validation for a C-band distributed RA with 4 pumps and 100-km standard single mode fiber demonstrates that the framework can provide highly--accurately designs of 3500 arbitrary gain profiles, in a controlled way, and predict the corresponding noise figure with an average maximum error of $\sim$0.3~dB.

\section{Inverse design with noise figure prediction}
\label{sec:Inverse_design}
In general, the pump power configuration defines the gain and the NF of a Raman amplifier. Therefore, the forward mapping of the RA can be described by $[\mathbf{G},\mathbf{NF}]=f(\mathbf{P})$, where $f(\cdot)$ is a differential equation operator~\cite{agrawalRaman}. $\mathbf{G} = [G_1, G_2, ..., G_{N}]^T$ and $\mathbf{NF} = [NF_1, NF_2, ..., NF_{N}]^T$ are discretized Raman gain and NF profiles over $N$ frequency channels. $\mathbf{P} = [P_1, P_2, ..., P_n]^T$ is the vector of pump power levels for $n$ pumps. 

\begin{figure}[t]
  \centering
  \includegraphics[width=0.44\textwidth]{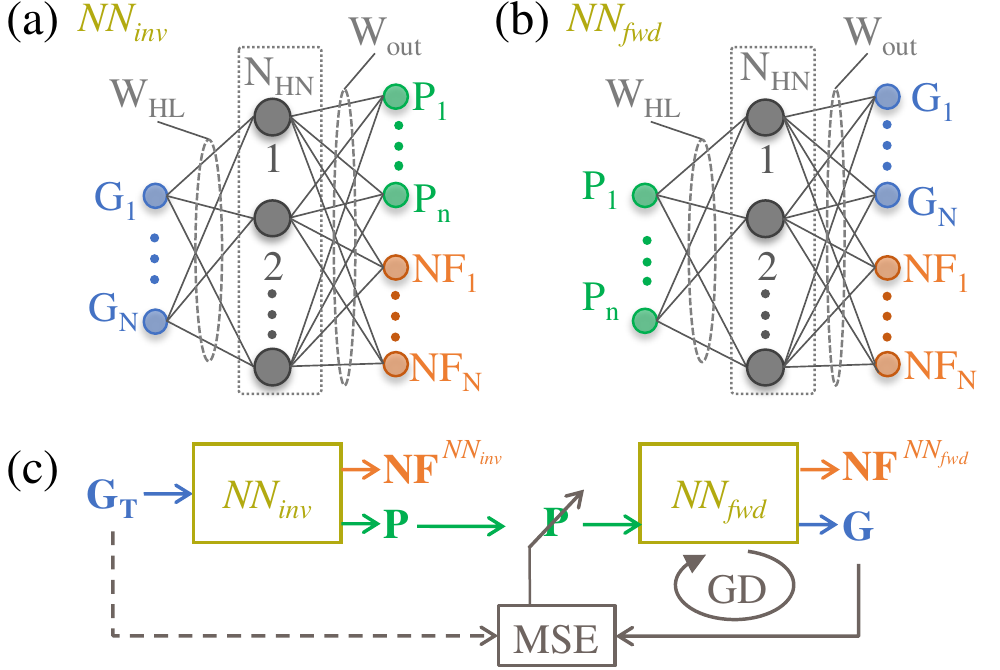}
\caption{Upgraded neural network models (with respect to~\cite{Zibar20}) for the (a) inverse ($NN_{inv}$) and (b) forward ($NN_{fwd}$) Raman amplifier mappings incorporating the noise figure prediction. (c) Full machine learning framework with the gradient descent (GD) routine used to fine--optimize the $NN_{inv}$ $\mathbf{P}$ predictions.}
\label{fig:NNmodels}
\end{figure}

To be able to simultaneously predict the pump power configuration and the NF for a target gain profile, we train an NN to learn the mapping between $\mathbf{G}$ and $[\mathbf{P},\mathbf{NF}]$. This NN is referred to as $NN_{inv}$ and is illustrated in Fig.~\ref{fig:NNmodels}(a). The sub-index $inv$ refers to the \emph{inverse design} provided by $NN_{inv}$, rather than the RA inverse mapping $f^{-1}(\cdot)$.
The RA input power is not considered as the robustness of the ML framework for different input signal power is discussed in~\cite{UiaraJLT_OFC20ext}.

The prediction accuracy of the $NN_{inv}$ can be improved by employing a fine--optimization technique presented in~\cite{Zibar20}. For that, a second ANN, referred to as $NN_{fwd}$, is employed to approximate the forward mapping $f(\cdot)$ as is illustrated in Fig.~\ref{fig:NNmodels}(b). The fine--optimization is illustrated in Fig.~\ref{fig:NNmodels}(c). It is based on a gradient descent (GD) routine that aims at finding the pump configuration that minimizes the mean squared error (MSE) between target and predicted gain profiles. The initial pump configuration is the one provided by $NN_{inv}$. Therefore, the GD algorithm requires only a few iterations to converge (around 37 on average) as the $NN_{inv}$ prediction is already quite accurate. Since the $NN_{inv}$ prediction takes only a few milliseconds, the total prediction time is defined by the fine optimization, being around 0.4 s on average (considering a standard computer setup, i.e., Intel(R) i7-8650U @1.90GHz, 16.0G RAM). During the GD iterations, $NN_{fwd}$ is used to provide fast gain predictions and gradient calculations for the gain MSE since it is fully differentiable. Note that the NF prediction is not optimized during the GD procedure. After the fine--optimization, $NN_{fwd}$ is applied one last time to compute an updated NF prediction considering the optimized pump power configuration as input.

Together with the optional fine--optimization step, $NN_{inv}$ comprises a comprehensive RA inverse design tool able to simultaneously provide pump powers and NF predictions for a given target gain profile $\mathbf{G_T}$, as illustrated in Fig.~\ref{fig:NNmodels}(c).

\section{Experimental setup and ANN training}
\label{sec:ExperimentalSetup}
\begin{figure}[t]
  \centering
  \includegraphics[width=0.44\textwidth]{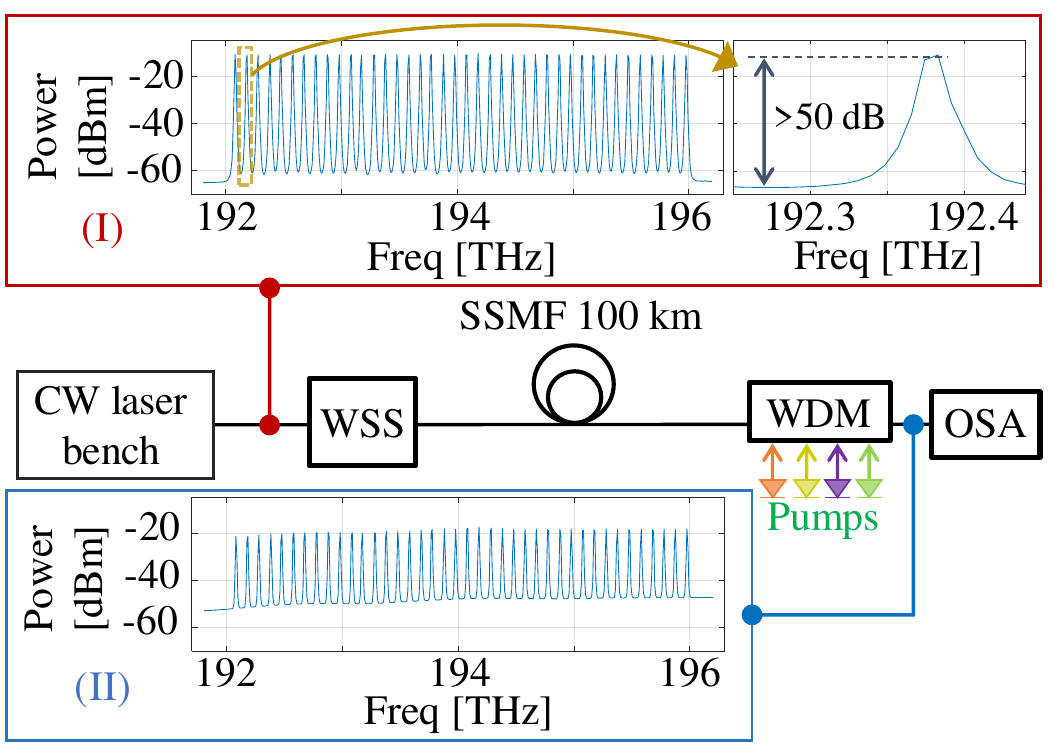}
\caption{Experimental setup for data--set generation showing an input and an output spectra to the Raman amplifier. The spectra are measured with the resolution bandwidth of 0.1~nm.}
\label{fig:setup}
\end{figure}

Fig.~\ref{fig:setup} shows the experimental setup used to generate the training and testing data--sets. It employs a distributed RA in a counter-propagating multi-pump configuration. The input signal is composed of $N$~=~40 continuous wave (CW) lasers, 100-GHz spaced on the ITU-T grid ranging from 192 to 196~THz. A wavelength selective switch (WSS) is used to flatten the input channels. The resulting spectrum at the input of the 100-km standard single-mode fiber (SSMF) is shown in Fig.~\ref{fig:setup}(I), where the ratio between the signal and noise power is highlighted. At the SSMF output, a commercial Raman amplifier pump module is used. It has $n$~=~4 pump lasers with fixed wavelengths $\bm{\lambda} = [1454.4, 1444.8, 1434.4, 1423.4]$~nm and adjustable powers for up to $\mathbf{P_{MAX}} = [145, 158.5, 180, 152.5]$~mW. The pumps are combined through a wavelength division multiplexing (WDM) coupler and an optical spectrum analyzer (OSA) captures the spectra after the RA.

The experimental data-set is generated as follows: $M$~=~7000 different pump power vectors $\mathbf{P}$ drawn from uniform distributions are applied to the experimental setup in Fig~\ref{fig:setup} and their respective output power spectra are measured. In this work, we consider the Raman on-off gain $\mathbf{G_{on-off}}$ calculated as the difference between the channel powers with the pumps turned on and off. The high input signals' OSNR levels (Fig~\ref{fig:setup}(I)) allow to consider the inputs nearly shot-noise-limited signals~\cite{BANEY2000122}. Therefore, the associated $\mathbf{NF}$ can be calculated using the measured amplified spontaneous emission (ASE) spectral density and the signal gain~\cite{BANEY2000122}. Here, the effective NF ($\mathbf{NF_{eff}}$), i.e. based on $\mathbf{G_{on-off}}$, is considered~\cite{agrawalRaman}. The data-set is given by $\mathcal{D}^{M\times (n+2N)}=\{\mathbf{X}_i,\mathbf{Y}_i|i=1,...,M\}$ with $\mathbf{X} = \mathbf{P}$ and $\mathbf{Y} = [\mathbf{G_{on-off}};\mathbf{NF_{eff}}]$ for $NN_{fwd}$, and $\mathbf{X} = \mathbf{G_{on-off}}$ and $\mathbf{Y} = [\mathbf{P};\mathbf{NF_{eff}}]$ for $NN_{inv}$. $\mathcal{D}$ is split into two equally-sized subsets: $\mathcal{D}_1$ (training+validation) and $\mathcal{D}_2$ (testing). The measurement errors for $\mathbf{NF_{eff}}$ and $\mathbf{G_{on-off}}$ estimation are below 0.1~dB.

$NN_{inv}$ and $NN_{fwd}$ are single--layer and fully connected NNs trained using random projection method~\cite{Huang11}. The model selection considered 10-fold cross-validation with 90\% of $\mathcal{D}_1$ used for training and 10\% for validation. It was used to obtain the following parameters: the standard deviation ($\sigma_{W_{HL}}$) of the normal distribution for the hidden layer's ($W_{HL}$) weights assignment, the regularization parameter ($\lambda$) for the regularized least squared error to calculate the output layer's weights ($W_{out}$), the number of hidden nodes ($N_{HN}$), and the hidden node activation function ($f_{act}$). This process results in $\sigma$ = 0.1, $N_{HN}$ = 1000 and $f_{act}(x)$ = $sin(x)$ for both NNs; and $\lambda$ = $10^{-8}$ and $10^{-2}$ for $NN_{inv}$ and $NN_{fwd}$, respectively. A linear activation function is considered for the output nodes ($f_{act}(x)$ = $x$) and model averaging with 20 NNs is employed just for the inverse model to make it less affected by the random $W_{HL}$ assignment. These NNs are trained in parallel, not significantly increasing the training complexity and time.

\begin{figure}[t]
  \centering
  \includegraphics[width=0.45\textwidth]{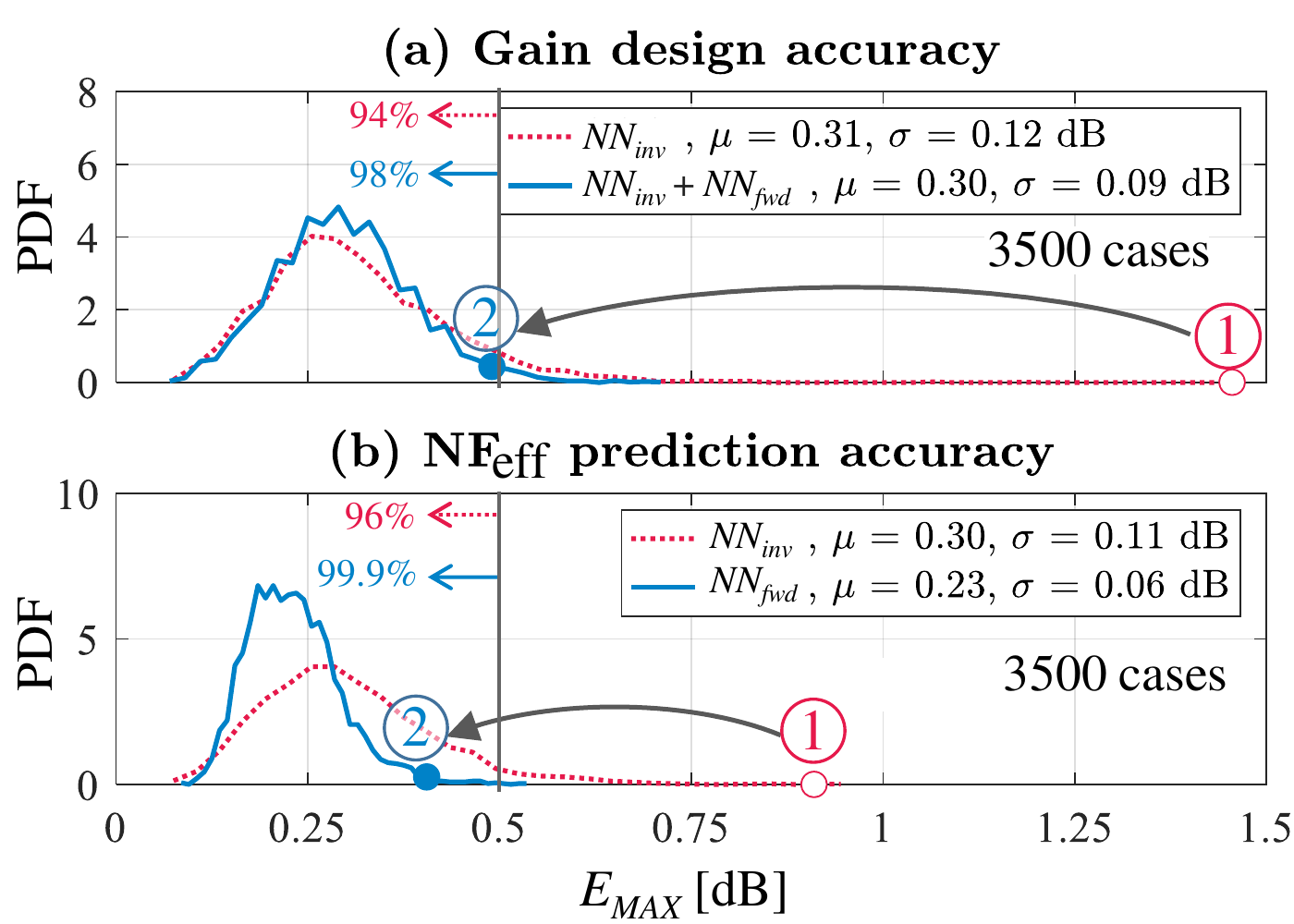}
\caption{Probability density function (PDF) of the maximum error ($E_{MAX}$) between (a) target and measured gains (from predicted pump powers) and (b) predicted and measured effective noise figure over the final test data-set $\mathcal{D}_2$, showing mean ($\mu$) and standard deviation ($\sigma$) values for the inverse designs ($NN_{inv}$) and fine--optimizations ($NN_{inv}+NN_{fwd}$).}
\label{fig:result_PDF}
\end{figure}

\section{Experimental validation results}
\label{sec:valid}
The accuracy of the framework to achieve the target gain profile and predict $\mathbf{NF_{eff}}$ is evaluated for: arbitrary gain profiles directly selected from the test data-set $\mathcal{D}_2$ and known to be achievable by the specific RA implementation, and flat gain profiles not part of the testing data-set. 

For the first test scenario, the $NN_{inv}$ is used to obtain the pump powers for 3500 arbitrary target gain profiles within the range 0 to 13.5~dB and estimate their respective $\mathbf{NF_{eff}}$. The corresponding pump powers returned by $NN_{inv}$ are then applied on the experimental setup (Fig.~\ref{fig:setup}) and new measurements are performed. The measured $\mathbf{G_{on-off}}$ profiles are then compared to the target $\mathbf{G_{on-off}}$ whereas the $\mathbf{NF_{eff}}$ measurements are compared with the predicted $\mathbf{NF_{eff}}$ returned by $NN_{inv}$. These comparisons are performed by computing the maximum absolute error $E_{MAX}$ over frequencies. The fine--optimization is also evaluated in a similar way. But in this case, measured $\mathbf{NF_{eff}}$ are compared with updated $\mathbf{NF_{eff}}$ predictions returned by $NN_{fwd}$, considering the fine optimized pump powers as input.

Fig.~\ref{fig:result_PDF} shows the probability density function (PDF) of $E_{MAX}$ for the gain designs and the $\mathbf{NF_{eff}}$ predictions. By just applying $NN_{inv}$ (red dashed lines), similar performances are achieved for both $\mathbf{G_{on-off}}$ design and $\mathbf{NF_{eff}}$ prediction in terms of $E_{MAX}$ mean $\mu$ and standard deviation $\sigma$ values. If the fine--optimization, referred to as $NN_{inv}$+$NN_{fwd}$ (blue solid lines), is considered, no improvement is observed for the gain design $E_{MAX}$ $\mu$ and $\sigma$. However, for the target gain case with the worst $NN_{inv}$ performance, the $E_{MAX}$ is reduced from $\sim$1.5 to $\sim$0.5~dB (point 1 to 2 in Fig.~\ref{fig:result_PDF}(a)). The overall worst performance for $NN_{inv}$+$NN_{fwd}$ is now $\sim$0.75~dB. Moreover, the number of cases with $E_{MAX}$ lower than 0.5~dB increases from 94 to 98\% when fine optimizing the pump powers. For the $\mathbf{NF_{eff}}$ predictions, both average and maximum $E_{MAX}$ values are improved after the fine--optimization, with the number of cases with $E_{MAX}$ lower than 0.5~dB increasing from 96 to 99.9\%, as shown in Fig.~\ref{fig:result_PDF}(b). However, the $NN_{inv}$ performance in predicting NF is still accurate, allowing an ultra fast pump and noise figure prediction for most cases where the inverse model alone is sufficient to provide accurate gain designs.

\begin{figure}[t]
  \centering
  \includegraphics[width=0.45\textwidth]{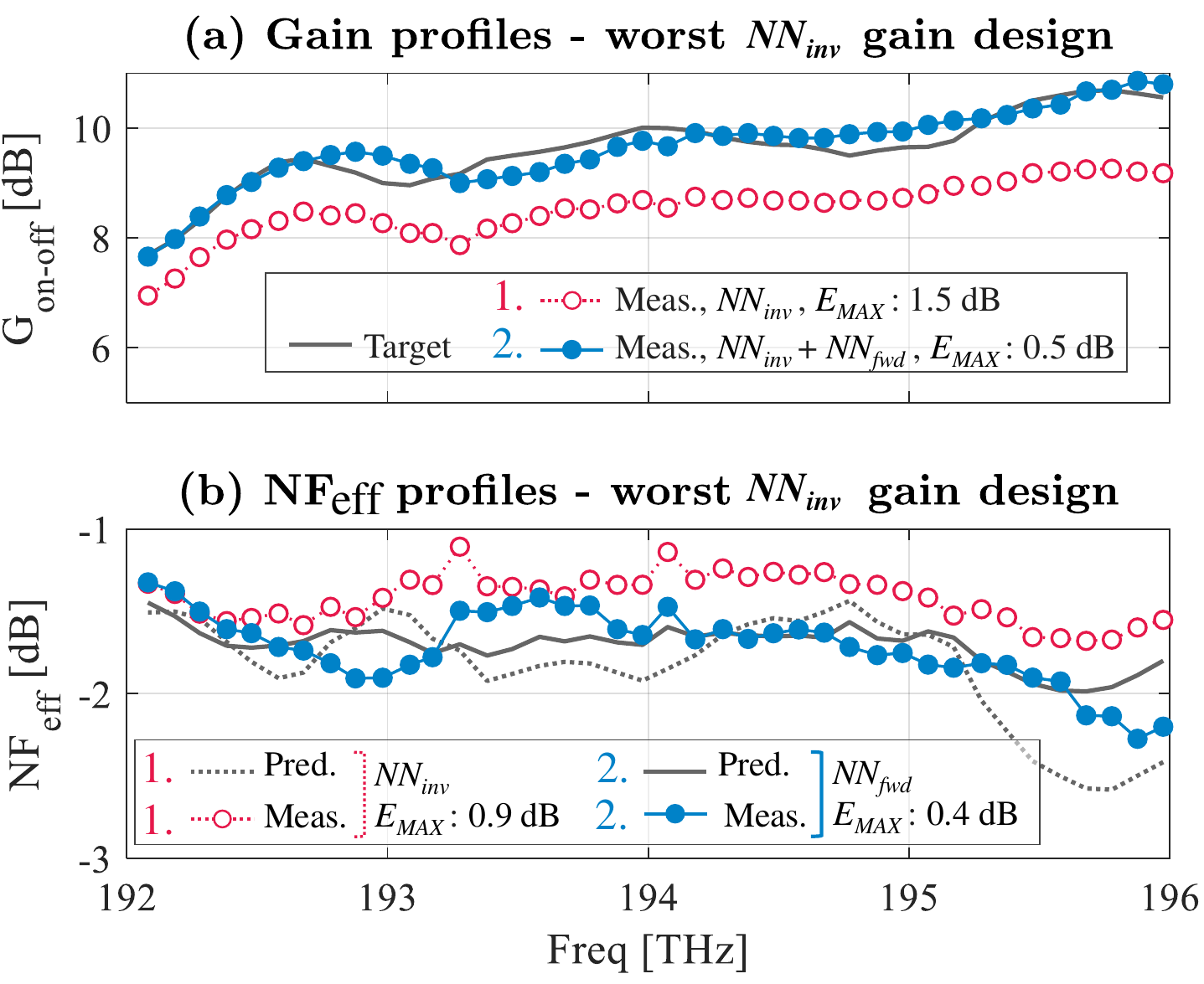}
\caption{(a) On-off gain and (b) effective noise figure spectral profiles for the worst case inverse design ($NN_{inv}$) and its corresponding fine--optimization ($NN_{inv}+NN_{fwd}$).}
\label{fig:result_curves}
\end{figure}

Fig.~\ref{fig:result_curves}(a) shows the target gain profile corresponding to the worst case (maximum $E_{MAX}$) when applying $NN_{inv}$. The measured on--off gain profiles for the design ($NN_{inv}$) and fine--optimization ($NN_{inv}+NN_{inv}$) are also shown. They correspond to points 1 and 2 in Fig.~\ref{fig:result_PDF}(a), respectively. The fine--optimization provides a gain profile significantly closer to the target gain, corresponding to $\sim$1~dB $E_{MAX}$ reduction when compared to the $NN_{inv}$ measured gain curve.

The worst $NN_{inv}$ gain design case also presents a high $\mathbf{NF_{eff}}$ prediction error (slightly lower than the worst $\mathbf{NF_{eff}}$ prediction), as indicated by point 1 in Fig.~\ref{fig:result_PDF}(b). Fig.~\ref{fig:result_curves}(b) shows the corresponding $NN_{inv}$ $\mathbf{NF_{eff}}$ prediction (gray dashed line) and measurement (red dashed line with empty circles). After fine--optimization, the updated $\mathbf{NF_{eff}}$ prediction (gray solid line), obtained by considering the optimized pump power values at the $NN_{fwd}$ input, and the new $\mathbf{NF_{eff}}$ measurement (blue solid line with circles) are closer to each other. This shows that the fine--optimization also improves the $\mathbf{NF_{eff}}$ prediction.

\begin{figure}[t]
  \centering
  \includegraphics[width=0.45\textwidth]{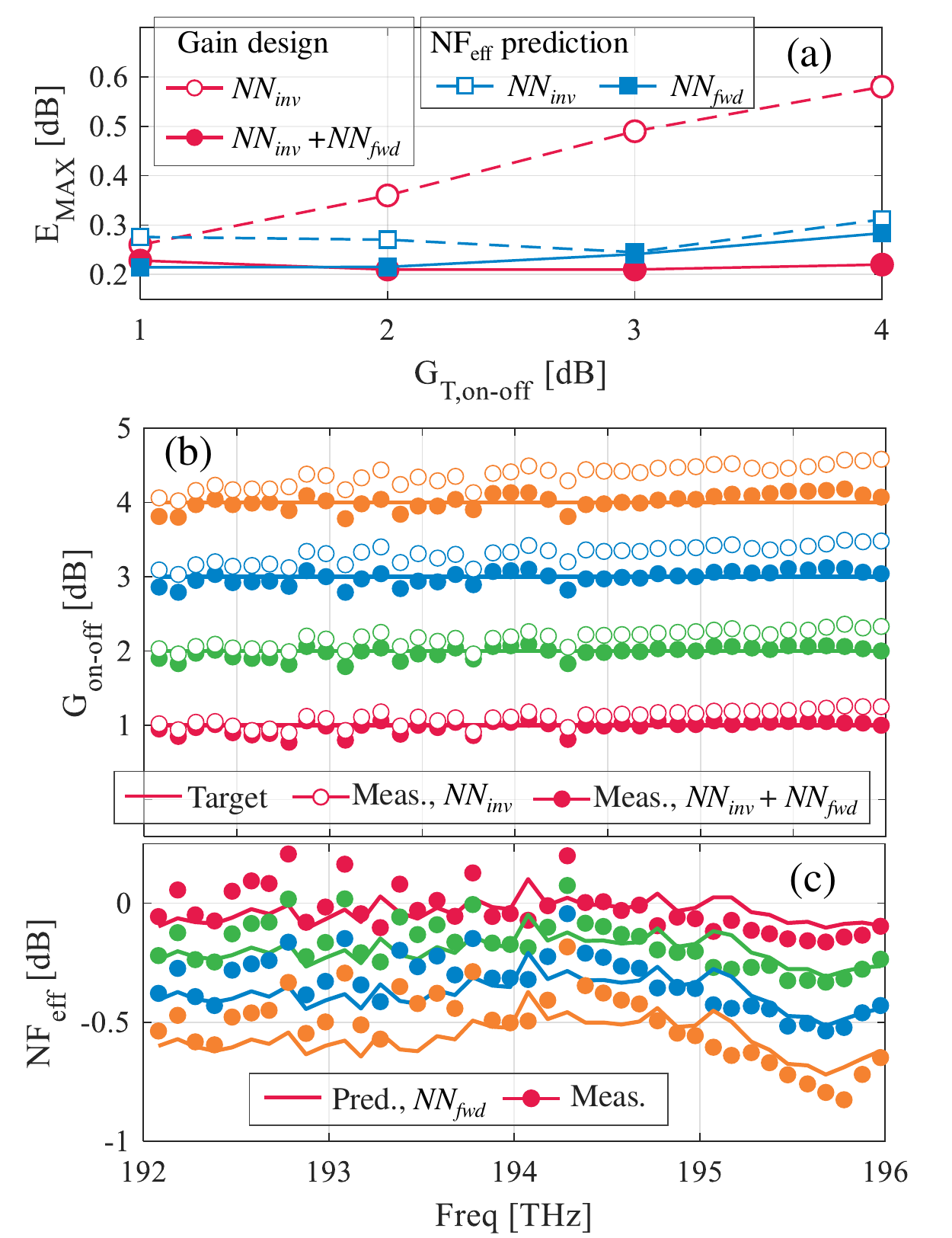}
\caption{(a) Maximum errors ($E_{MAX}$) for the gain designs and the effective noise figure predictions for different target on--off gain cases. The gain designs consider $NN_{inv}$ (inverse design) and $NN_{inv}+NN_{fwd}$ (fine--optimization), whereas $\mathbf{NF_{eff}}$ is predicted by $NN_{inv}$ (inverse design) and updated by $NN_{fwd}$ (after fine optimize the gain design). Comparing the profiles for (b) on--off gain (target and measured) and (c) $\mathbf{NF_{eff}}$ (predicted and measured), the latter considering only $NN_{fwd}$ prediction.}
\label{fig:result_flat}
\end{figure}

As second test scenario, we investigated the design of flat gain profiles and their respective $\mathbf{NF_{eff}}$ estimation. This analysis considers four flat target gains, from 1 to 4~dB. Flat gains higher than 4~dB cannot be achieved due to pump power limitations for the considered setup~\cite{UiaraJLT_OFC20ext}. The same validation procedure carried for the arbitrary gains was performed for the design ($NN_{inv}$) and fine--optimization ($NN_{inv}+NN_{fwd}$). The obtained results are shown in Fig.~\ref{fig:result_flat}. Fig.~\ref{fig:result_flat}(a) shows $E_{MAX}$ as a function of the evaluated target gain for the gain designs and $\mathbf{NF_{eff}}$ predictions. The gain design accuracy using $NN_{inv}$ decreases with the target gain. The fine--optimization reduces the gain design $E_{MAX}$, keeping it $<$0.3~dB for all evaluated target gains. The $\mathbf{NF_{eff}}$ prediction performance is almost constant over the evaluated gain, with no improvements after fine optimizing the pumps and considering the new $NN_{fwd}$ $\mathbf{NF_{eff}}$ predictions. 

This is the opposite behavior observed for the arbitrary gains. For the flat gains, the gain design accuracy is improved by the fine--optimization, while the $\mathbf{NF_{eff}}$ prediction does not change significantly. However, notice that the worse performance in achieving flat gains is $\sim$0.5~dB. This value is already accurate in terms of gain design. Therefore, since target and measured gains are close to each other for $NN_{inv}$ and $NN_{inv}+NN_{fwd}$, the $\mathbf{NF_{eff}}$ predictions will also be similar. 

Fig.~\ref{fig:result_flat}(b) shows targeted and predicted flat gain profiles. For the flat gain profile design, the fine--optimization reduces the gain mismatch especially for higher frequencies and gains. The $\mathbf{NF_{eff}}$ profiles are shown in Fig.~\ref{fig:result_flat}(c), where just $NN_{fwd}$ predictions and the measurements with finely optimized pumps are shown. The fine--optimization has a better $\mathbf{NF_{eff}}$ prediction performance for higher frequency channels since it reduces the gain mismatches between targets and measurements in this region. The errors in the low-frequency region are related to the channel gain variations. Comparing Fig.~\ref{fig:result_flat} (b) and (c), notice how changing the average gain of the amplifier can significantly affect the  $\mathbf{NF_{eff}}$. This demonstrates how important it is to properly identify these $\mathbf{NF_{eff}}$ changes and their consequences on the overall signal performance before applying any RA gain adjustments.

\section{Conclusions}
We present and experimentally validate a comprehensive machine learning framework that offers a comprehensive design and performance characterization tool for the Raman amplifier. The framework provides accurate pump power configuration and noise figure predictions, both with 0.3~dB of averaged maximum error over more than 3500 target gains profiles, including flat and arbitrary shapes. The ability to shape the gain profiles is an effective way to compensate for wavelength-dependent gain/loss in optical communication systems. Therefore, accurate prediction of the impact of such gain profile adjustments on the noise performance of the optical amplifier is critical to ensure reliable transmissions in future high-capacity optical networks.

\vspace{0.2cm} 
\noindent\textbf{Funding.} This project has received funding from the European Research Council through the ERC-CoG FRECOM project (grant agreement no. 771878), the European Union's Horizon 2020 research and innovation programme under the Marie Sk\l{}odowska-Curie grant agreement No 754462, the Villum Foundations (VYI OPTIC-AI grant no. 29344), and Ministero dell’Istruzione, dell’Università e della Ricerca (PRIN 2017, project FIRST).

\vspace{0.1cm}  
\noindent\textbf{Acknowledgments.} We thank Karsten Rottwitt for the fruitful discussions about the noise figure of Raman amplifiers.

\vspace{0.1cm}  
\noindent\textbf{Disclosures.} The authors declare no conflicts of interest.

\bibliography{sample}

\end{document}